\documentclass[12pt]{article}
\usepackage{vmargin}
\usepackage{pa}
\usepackage{amsmath}
\usepackage{amsxtra}
\usepackage{amsbsy}
\usepackage{amssymb}
\usepackage{indentfirst}
\usepackage{array}
\usepackage{dcolumn}
\usepackage{bm}
\usepackage{booktabs}
\usepackage[pdftex]{graphicx}
\usepackage{xcolor,colortbl}
\definecolor{lg}{gray}{.8}
\definecolor{dg}{gray}{0.6}
\usepackage{subcaption}
\usepackage[pdftex=true,%
            pdftoolbar=true,%
            pdfmenubar=true,%
            pdfauthor = {Nathaniel\ Beck},%
            pdfcreator = {PDFLaTeX},%
            pdftitle = {Binary DVs},%
            colorlinks=true,%
            urlcolor=blue,%
            linkcolor=blue,%
            citecolor=blue,%
            implicit=true,%
            hypertexnames=false%
            ]{hyperref}
 \usepackage{pdfsync}
\usepackage{url}
\usepackage{xspace}
\usepackage{numprint}
\usepackage{longtable}
\newcommand{\newc}{\newcommand}
\newc{\skipc}[1]{\multicolumn{#1}{c}{}}
\newc{\N}{\hphantom{0}}
\newc{\M}{\hphantom{$-$}}
\newc{\barr}{\begin{array}}
\newc{\earr}{\end{array}}
\newc{\bcen}{\begin{center}}
\newc{\ecen}{\end{center}}
\newc{\subsect}{\subsection*}
\newc{\beqann}{\begin{eqnarray*}}
\newc{\eeqann}{\end{eqnarray*}}
\newc{\beqa}{\begin{eqnarray}}
\newc{\eeqa}{\end{eqnarray}}
\newc{\beqnn}{\begin{displaymath}}
\newc{\eeqnn}{\end{displaymath}}
\newc{\beq}{\begin{equation}}
\newc{\eeq}{\end{equation}}
\newc{\tabst}{\begin{table}\centering}
\newc{\tabstp}{\vfill\begin{table}[p]\centering}
\newc{\tabend}{\end{tabular}\end{table}}
\newc{\btab}{\begin{tabular}}
\newc{\fn}{\footnote}
\newc{\cols}{\multicolumn}
\newc{\vm[1]}{\bm{\mathrm{#1}}}
\newc{\ea}{et al.}

\newcommand{\bb}{\ensuremath{\pmb{\beta}}\xspace}

\newcommand{\gb}{\ensuremath{\beta}\xspace}

\newcommand{\bX}{\ensuremath{\pmb{X}}\xspace}

\newcommand{\bbh}{\ensuremath{\hat{\pmb{\beta}}}\xspace}
\newc{\tsl}{\textsl}
\newc{\bit}{\begin{itemize}}
\newc{\eit}{\end{itemize}}
\newcommand{\bxgi}{\ensuremath{\pmb{x}_{g,i}\xspace}}
\newcommand{\ygi}{\ensuremath{y_{g,i}\xspace}}

\newcolumntype{.}{D{.}{.}{-1}}
\newcolumntype{d}[1]{D{.}{.}{#1}}
\newcolumntype{E}{D{.}{.}{3}}

\setpapersize{USletter}
\setmarginsrb{1in}{1in}{1in}{1in}{0pt}{0mm}{0pt}{0.25in}

      \title{Estimating  grouped data  models with a binary dependent
  variable and fixed effect via logit vs OLS: the impact of dropped units}
\author{Nathaniel Beck\thanks{Department of Politics;
           New York University;  New York, NY 10003 USA;
           \texttt{nathaniel.beck@nyu.edu}.
     }
       \date{\today}}

\begin{document}

\maketitle
\pagebreak

\begin{abstract}
This letter deals with a very simple issue: if we have
grouped data with a binary dependent variable and want to include
fixed effects (group specific intercepts)  in
the specification, is Ordinary Least Squares (OLS)  in any way
superior to a  logit form because the OLS method \emph{appears} to
keep all observations whereas the logit drops all groups which have
either all zeros or all ones on the dependent variable? It is shown that OLS averages the estimates for the all zero (and all one)
groups, which by definition have all slope coefficients of zero, with
the slope coefficients for the groups with a mix of zeros and
ones. Thus the correct comparison of OLS to logit is to only look at
groups with some variation in the dependent variable. Researchers
using OLS are urged to report results both for all groups and for the
subset of 
groups where the dependent variable varies. The interpretation of the
difference between these two 
results depends upon assumptions which cannot be empirically assessed.
\end{abstract}
\newpage
\section{Introduction}
\label{sec:intro}
Many\ applied researchers include ``fixed effects'' (unit specific intercepts) to account for unmodeled
heterogeneity in grouped data  analyses; these fixed effects lead to
interesting issues.
This is a well worked area when the dependent
variable is continuous (Greene 2018, ch. 11.4).  The situation is more
complicated when the dependent variable is binary, though again the
theory is well worked out (Greene 2018, ch. 17.7.3). In
particular, the group mean centering solution for 
estimating a model with fixed effects and a continuous
dependent variable does not carry over to non-linear models, such as
logit. Obviously a standard logit model with fixed effects
(``LOGITFE'')  \emph{can} be
estimated; one just adjoins the unit specific dummy variables to the
specification. The question is whether one should estimate LOGITFE as
compared to rival estimators.

The famous work of  results of Neyman and Scott (1948)  showed that, as the
number of incidental parameters goes to infinity along with    the sample size, simply including these incidental parameters leads to biased
estimates. In the grouped data case the number of incidental
parameters is the number of group specific intercepts. This has put
off researchers from using LOGITFE. This became more pronounced when 
Chamberlain (1980)
proposed the conditional logit (``CLOGIT'')  model, which
conditions out the fixed effects by conditioning on the number of
successes (1's) and failures (0's) in the group. This conditional
approach is what Neyman and Scott proposed in general to solve the
incidental parameters problem. Thus CLOGIT produces unbiased
estimators in the presence of group specific intercepts. 

The gains from using
the CLOGIT  hinge  on the type of data that is being studied, viz.,
are the groups large or  small. Thus we might
have groups of size two in a two wave ``panel'' study; at the other
extreme, we often see ``time-series--cross-section-data'' with well
over 30 observations per unit.
It is now well known  that LOGITFE is essentially unbiased  if group size is
large enough; simulations show that  large enough is say 20 or more
(Katz 2001; Greene 2004; Coup\'{e} 2005) observations per group.
Beck (2018) has recently shown that in such circumstances LOGITFE
has similar mean squared error loss to CLOGIT. Thus for data where
group sizes are large there is no advantage of CLOGIT over LOGITFE.

CLOGIT also comes with costs; since the
fixed effects are conditioned out rather than estimated, marginal
effects of covariates cannot be estimated with CLOGIT. These days
marginal effects are \emph{de rigor}; LOGITFE easily allows for the
estimation of marginal effects since the fixed effects are estimated. Thus where group sizes are large
(say over 20 or so), LOGITFE should be used instead of CLOGIT.\fn{This
  is not to say that it always or usually is; estimators with even
  small amounts of bias are often scorned even if they have other
  advantages. This is not the place to fight this battle even if the
  outcome should be obvious.} Since this
is the type of data considered in the list letter, I only discuss
LOGITFE.\fn{A search of the most recent five years of political
  science/international relations articles in JSTOR found none using
  CLOGIT for the very small group size case; about two or three per
  year use it in the large group case. The search did not turn up any
  articles using LOGITFE. Most articles with binary dependent
  variables started with the OLSFE specification.}

FELOGIT \emph{appears} problematic in that it
drop all groups from estimation that are either all failures and/or
all successes 
(``ALLZERO'').  In data where successes are rare (say wars or
coups) this can lead to the loss of much of the data.\fn{The largest number of cases dropped
 due  to  groups with no successes that I know is is the Green, Kim
 and Yoon (2001)  fixed effects analysis of
  Militarized Interstate Disputes, where 93\% of the data does not
  enter the likelihood function.} The reason why ALLZERO
   groups are dropped is well laid out in the standard
  econometric texts already cited. The simple reason is that the fixed
effects do a perfect job of predicting outcomes in ALLZERO groups, with
the group specific intercept estimator for those groups converging to negative
infinity. Similarly, for all success groups the group specific
intercept estimator converges to positive infinity and so these groups
are dropped from estimation identically to all failure groups;  the
values of the covariates for either 
those groups  have no effect on the likelihood function. To
simplify exposition  I focus on all failure groups only, but this
leads to  no
loss of generality; ALLZERO groups include  groups with only failure
\emph{and} groups with 
only success. 

This has led to a somewhat
paradoxical situation, with applied researchers choosing to use a
linear probability model with unit specifici intercepts
(``OLSFE''). OLSFE does not drop all failure or all success units and
so appears to work with all the data.\fn{There may be other reasons
  why OLSFE is common, such as it is easy to read off the marginal
  effects, which are just the estimated parameters; for LOGITFE there
  is an additional (trivial) step. Casual conversation with those using OLSFE
  believe that the results are the same as with LOGITFE, and are
  obtained at a lower (intellectual) cost. This letter is dealing with
  the question of whether the two methods are as similar as
  believed.} I use the word paradoxical because logit is surely the model
of choice with cross-sectional binary dependent variable data, but
OLSFE seems to have that status with grouped binary dependent variable
data.

The
argument for LOGITFE vs OLSFE at least partly turns on what appears to
be the loss of the ALLZERO groups  in LOGITFE, which can radically
change the properties of the data set. The purpose of this letter is
to elucidate this  effect. This letter deals
only with grouped data where group size is large enough (say 20 or
more) so that
LOGITFE produces essentially unbiased and efficient estimators,
allowing me to only compare LOGITFE and OLSFE.

Researchers are aware of this issue, but they often seem to take as
either a robustness issue or as a reason to pursue a non-obvious line
of attack. This can be seen in two recent examples in very prominent
journals. Wright, Frantz and Geddes (2103) examines the link between oil wealth and
autocratic regime survival using a country-year design with a binary
dependent variable denoting whether a regime survived a given year.
The article notes that fixed effects are generally included in models
similar to theirs in order to deal with unobserved unit
heterogeneity. Wright, Frantz and Geddes (2103, p. 294) goes on to say 
``[T]his strategy, however, drops [between 26 and 64] countries from
the analysis that do not experience [various types of]  regime
change.... Dropping countries that do not experience regime change may
bias estimates downward by selecting only those where regime change
has occurred in the sample period, particularly if those stable
political systems have high oil wealth. Below, we investigate the
possibility that this restriction on the sample induces selection bias
[by dropping fixed effects from the logit model].''The results differ a
lot but is that due to the change in sample or the dropping of fixed
effects; if unobserved unit heterogeneity is important, dropping fixed
effects from the model does not seem ideal.

In a second important example, 
 Besley and Reynal-Querol (2011)  studies whether democracies provides
more educated leaders by estimating models, for example, of the
probability of a leader having a graduate degree in a large number of
countries, where the specification includes country fixed
effects. This article provides both OLS estimates of the LPMFE
specification and CLOGIT estimates of the LOGITFE
specification. Besley and Reynal-Querol (2011, p. 559)  estimates  the LMPFE specification (without
justification) and then only mentions using CLOGIT (LOGITFE) as a robustness check.  
``[In the CLOGIT (LOGITFE)
specification], we estimate a conditional logit
model to recognize the discrete nature of the lefthand-side
variable. The core finding of [the LPMFE model] remains.''  This is
clearly correct if we only care about the sign and significance of a
coefficient, but, as we shall see, the difference between the two
estimations is not trivial, albeit not enormous.

To keep
this clear we need a bit of notation. Section~\ref{s:all0} then shows
the consequence of including the ALLZERO groups in the OLS estimation and
the difference between OLSFE and LOGITFE as a function of dropping
those group. Section~\ref{s:examples} reanalyzes one 
Besley and Reynal-Querol (2011) result to show the practical import of the the
analytic results. 
The conclusion discusses some
interpretative issues in these differences and indicates when one
might prefer the logit or OLS estimates of the marginal effects.

\section{Notation} \label{s:notation}

Let  \ygi\ be a binary dependent variable with the exogenous
covariates being
\bxgi, where 
 $g$ indexes groups and $i$ indexes particular units in a group.
  It simplifies notation to assume that all groups are of the
same size, and dropping this one extra subscript has no consequences
for the argument:  let this be group size be $N$, with $G$  being  the
number of groups. Let the number of covariates be $k$. $\alpha_i$ refers to the fixed effect for group
$g$, that is the group specific intercept.

 What is critical for this article is that $G$ is
fixed; 
asymptotics are in terms of $N$. This surely holds for a very common
type of data seen in comparative politics: time-series--cross-section
data. Here $G$ is the number of countries of other units being
compared, and $N$ (often denoted $T$) is the number of periods (usually years) that a unit
is observed. Note that $G$ may be large, but it is fixed; even if we
compare, say, 5000 US counties, the number of counties is not
growing
infinitely large. In this data $N$ is often (but not) always
reasonably large (say 20 or more). What is critical is that in
asymptotic thought experiments $\lim_{N\rightarrow \infty} \frac{G}{N} =
  0$ and that there are no asymptotics in a fixed (perhaps large)
  $G$. But the time-series-cross-section structure of the data is not
  at all critical here; what is critical is that asymptotics are in
  $N$ and not $G$ and these asymptotics yield the stated limit. This
  would hold if people are divided into ethnic groups with the number
  of people studied in each group being sufficiently large. It would
  \emph{not} hold for, say, a two or short wave panel, nor would it
  hold for educational studies where the group is a classroom. 

  The LOGITFE model is
  \begin{equation}
P(\ygi=1) = \frac{1}{1+e^{-(\bxgi\bb +\alpha_g)}}
\end{equation}
where $\alpha_i$ is the group specific intercept. 
.The OLSFE model is
\begin{equation} 
\ygi = \bxgi\bb + \alpha_g+ \epsilon_{g,i} \label{eq:linear}
\end{equation}
and $P(\ygi=1)$ is estimated in the obvious way.
For both OLS and logit we can then estimate $\frac{\partial P}{\partial
  \bxgi}$, the marginal effect of the covariates on $P(y=1$).

There may be some groups where every member of the group has
$y=0$ (``ALLZERO'' groups).\fn{Groups with only successes or a mixture
  of ALLZERO and all success groups yield identical results as shall
  be shown in the next section. I therefore refer to ALLZERO groups as
  including all success groups, without loss of generality. While
  groups with all failures seem more common in political science,
  there is no reason we cannot observe a mis of all failure and all
  success groups.} A common reason for having groups with all zeros
is that events ($y=1$) may be rare; think of whether a country has a
civil war in a year, with many countries having no civil war
ever. This letter  deals with the issue of the consequences of such
groups, and it is shown that that the consequences, not surprisingly,
increase as the number of ALLZERO  groups increases.

\section{Differences between what is estimated with LPMFE and
  LOGITFE} \label{s:all0}

As noted, any of the methods used to estimate a LOGITFE specification
drop  the ALLZERO groups. The LPMFE model estimated by OLS does use
information on all the groups. To see the consequences of this, note
that the OLS estimate of \bb is a weighted average of the estimates in
the ALLZERO and the other (``NOTALLZERO'')  groups.\fn{Any estimator can be
  seen as a combination of estimators for subgroups of data; this is a
  long standing idea in econometrics, with perhaps the best known and
  long standing example being the the Chow (1960) test of the equality of
  regression lines in two subsets of data. The
  calculations here are even simpler because of the nature of the
  dependent variable in the ALLZERO group.}
For the ALLZERO groups,
$\ygi=0$ so the OLS estimate of \bb for those data is zero.\fn{All group intercepts in
the ALLZERO group are also zero, and the fit appears to be
perfect, with the estimated group specific intercepts being zero.  It is necessary to assume some variation in the covariates
within 
the ALLZERO groups so the model is identified. Note that for groups
with only successes, the estimate of \bb is still zero, and the fit is
still perfect, although the estimated intercepts are now one. That is
why it is not necessary to separate all failure and all success
groups.} Thus the OLS estimate of \bb in the full data is a weighted
average of the OLS estimate of \bb in the NOTALLZERO groups and
$mathbf{0}$ where the weights depend non-linearly  on the covariates in the two
groups.

For the OLS computations, it is simplest to work with group
mean centered data to avoid putting the group intercepts in the
specification.  Let $\pmb{\tilde{\mathbf{X}}}$ and $\pmb{\tilde{\mathbf{y}}}$ be the group mean centered
data and let 
 $\pmb{\tilde{\mathbf{X}}_0}$ be the covariate matrix for the ALLZERO
groups with $\pmb{\tilde{\mathbf{X}}}_1$ being the corresponding matrix for the
NOTALLZERO groups with   $\pmb{\tilde{\mathbf{y}}}_1$ being  the
group mean centered  vector of
observations on $y$ for the NOTALLZERO (and obviously the
corresponding vector for the ALLZERO groups is
$\pmb{\tilde{\mathbf{y}}}_0=\mathbf{0}$. The subscript $01$ refers to
the complete data.

Thus the OLS estimate of $\bb_{01}$
for the entire data set   are given by 
\begin{equation}
\bbh_{01} =
\pmb{(\tilde{\mathbf{X}}_1'\tilde{\mathbf{X}}_1+\tilde{\mathbf{X}_0}'\tilde{\mathbf{X}}_0)^{-1}(\tilde{\mathbf{X}}_1'
\tilde{\mathbf{y}}_1)}
\label{eq:b01}
\end{equation}
whereas the corresponding estimate for the NOTALLZERO 
groups ($\bb_1$)
is given by 
\begin{equation}
\bbh_1=
\pmb{(\tilde{\mathbf{X}}_1'\tilde{\mathbf{X}}_1)^{-1}(\tilde{\mathbf{X}}_1'
\tilde{\mathbf{y}}_1)}.
\label{eq:b1}
\end{equation} 
We can also compare the variance covariance matrix of the two
estimates. For the entire data set this matrix is  
\begin{equation}
\pmb{(\tilde{\mathbf{X}}_1'\tilde{\mathbf{X}}_1+\tilde{\mathbf{X}}_0'\tilde{\mathbf{X}}_0)^{-1}}\widehat{\sigma_{01}^2}
\label{eq:s01}
\end{equation}
whereas the corresponding estimate for the NOTALLZERO 
groups ($\bb_1$)
  is given by 
\begin{equation}
\pmb{(\tilde{\mathbf{X}}_1'\tilde{\mathbf{X}}_1)^{-1}}\widehat{\sigma_{1}^2}
\label{eq:s1}
\end{equation}
where $\widehat{\sigma_{01}^2}$ and $\widehat{\sigma_{1}^2}$ refer to
estimates of the standard error of the regression in the full and
restricted data sets respectively.

It is immediately obvious that the two equations only differ by the 
$\pmb{\tilde{\mathbf{X}}_0'\tilde{\mathbf{X}}}_0$ portion of the $\bX'\bX$ matrix that is being
inverted. Alternatively, it is obvious that the OLS estimates for all
the data is a weighted average of $\mathbf{0}$ and the $\bbh_1$;  $\bb_{01}$ shrinks
$\bb_1$ towards $\mathbf{0}$.  The amount of shrinkage is a 
 somewhat complicated function that depends on the
relative scale of $\pmb{\tilde{\mathbf{X}}_0'\tilde{\mathbf{X}}_0}$ and
$\pmb{\tilde{\mathbf{X}}_1'\tilde{\mathbf{X}}_1}$.  As the proportion
of ALLZERO groups goes up, $\bbh_{01}$ goes to $\mathbf{0}$, but the path
may not always be monotonic for all components of $\bbh_{01}$.

The variance covariance matrix of the estimates has two components
which move in different directions as we move from the entire data set
to the NOTALLZERO data set. The estimated $\sigma^2$ will get smaller,
since we are eliminating non-homogenous cases; however 
the $\pmb{\tilde{\mathbf{X}}_1'\mathbf{\tilde{X}}_1}$ matrix in the NOTALLZERO data
 will also be smaller
in scale than the corresponding
 $\pmb{\tilde{\mathbf{X}}_{01}'\mathbf{\tilde{X}}_{01}}$ 
matrix used to estimate
the variance covariance matrix of of $\bbh_{01}$. Note however that the estimated standard error of the
regression will be limited in how much it changes since the variance
of $\pmb{\tilde{\mathbf{y}}}$ is limited by it being a binary variable; the $\pmb{\tilde{\mathbf{X}}'\tilde{\mathbf{X}}}$ matrix is
not similarly limited by any scaling, and so could shrink considerably
as the ALLZERO cases are dropped. 
 Usually , the estimated standard errors of $\bbh_{1}$ will be
 smaller than the corresponding estimates for $\bbh_{01}$ The
change in \bbh\  and its estimated standard error
offset, and so we usually see smaller impacts of dropping the ALLZERO
groups  on the $t$-ratio
associated with $\bb_{01}$ as compared to $\bb_1$. This smaller change
in $t$-ratio  may be one reason
that authors are content to conclude that the substantive results from
LOGITFE  are similar to those of LPMFE. But we should go beyond simply
inquiring as to the sign of a coefficient and whether its
``significance'' is beyond some standard threshold to actually looking
at coefficients.

It is very simple to see what is going on by looking at the scalar $x$
case, where once again $\tilde{y}$ and $\tilde{x}$ have been group mean centered. The OLS estimate of $\gb_{01}$ 
for the entire data set is given by 
\begin{equation}
\hat{\beta}_{01} =
\frac{\sum\limits_{\text{NOTALLZERO}} \tilde{x}_{g,i}\tilde{y}_{g,i}}{\sum\limits_{\text{ALLDATA}} \tilde{x}_{g,i}^2} 
\label{eq:sb01}
\end{equation}
whereas the corresponding estimate for the NOTALLZERO 
groups ($\gb_1$)
  is given by 
\begin{equation}
\hat{\beta}_{1} =
\frac{\sum\limits_{\text{NOTALLZERO}} \tilde{x}_{g,i}\tilde{y}_{g,i}}
{\sum\limits_{\text{NOTALLZERO}} \tilde{x}_{g,i}^2}.
\label{eq:sb1}
\end{equation}
These  two equations  differ only  by an extra $\sum\limits_{\text{ALLZERO}}
\tilde{x}_{g,i}^2$ in the denominator of Equation~\ref{eq:sb01}; this extra
term is the sum of squares and so non-negative, so $\hat{\gb}_{01} < \hat{\gb}_1$. 
The standard error for $\hat{\gb}_{01}$ for the entire data set is  given by
\begin{equation}
\sqrt{\frac{\widehat{\sigma_{01}^2}}{\sum\limits_{\text{All Data}} \tilde{x}_{g,i}^2}}
\label{eq:ss01}
\end{equation}
whereas the corresponding standard error  for the NOTALLZERO 
groups ($\hat{\gb_1}$)
  is given by 
\begin{equation}
\sqrt{\frac{\widehat{\sigma_{1}^2}}{\sum\limits_{\text{NOTALLZERO}} \tilde{x}_{g,i}^2}}
\label{eq:ss1}
\end{equation}
where again the extra summation terms in the denominator must be positive.

For the scalar case it is obvious that including the ALLO groups 
shrinks $\hat{\gb_1}$ towards zero, where the amount of shrinkage
depends on how many ALLZERO groups there are and the variation of the
centered $x$'s in those groups. The estimated standard error of $\gb_1$
also gets smaller (in general), since the larger denominator due to
$\sum\limits_{\text{ALLZERO}} \tilde{x}_{g,i}^2$ will almost always offset the increase
in the estimate of the standard error of the regression due to the
greater heterogeneity of y of the full data set. This again leads to
offsetting effects in changing $t$-ratios. 

\section{Examples} \label{s:examples}

If readers need convincing of the mathematics, one example should
do. Here I reanalyze the Besley and Reynal-Querol (2011)  results cited previously since the article
is important and the replication data were provided by the author. 
It is easy to compare the LPMFE and LOGITFE results of the two
 estimate for the effect of democracy
on whether a leader had a graduate degrees. These results are
presented in Table 1 of the original article, with Column 1 being the LPMFE
model and Column 3 being the LOGITFE model. The regression results are
based on 1146 country-year observations; the logit results lose 190
of 
those because in some countries no leader ever had a graduate
degree. I present in Table 1 re-analyses of a slightly simpler specification
here, so that this letter can focus on the dropped cases issues and
use LOGITFE instead of CLOGIT; when comparable, the results here are
similar to those of the original article.

\setlength{\tabcolsep}{2pt}
\begin{table}[tb]
 \begin{center}
   \begin{tabular}{l.......}
     \skipc{1} & \cols{2}{c}{OLSFE/All} &
                                                   \cols{2}{c}{OLSFE/No
                                        
                                                   ALL0 } &
                                                            \cols{2}{c}{LOGITFE}
     & \cols{1}{c}{OLSFE/ALL0 only } \\
 \cline{2-8} 
 \skipc{1} & \cols{1}{c}{$\hat{\beta}$}  & \cols{1}{c}{SE} &
                                                             \cols{1}{c}{$\hat{\beta}$}
           & \cols{1}{c}{SE}  & \cols{1}{c}{$\hat{\beta}$}  &
                                                              \cols{1}{c}{SE} & \cols{1}{c}{$\hat{\beta}$}\\
Democracy &  0.260 & 0.043 & 0.295 & 0.046  & 1.748 & 0.269 & 0.000\\
     Av. Marg. Effect (Dem) & \skipc{4} & 0.297 & 0.040  \\
     N & \cols{2}{c}{1146} & \cols{2}{c}{956} & \cols{2}{c}{956} & \cols{1}{c}{190}\\
  \end{tabular}
 \caption{LPMFE and LOGITFE estimates of effect of a democracy dummy
   variable on probability a leader has  a graduate degree in a given
   country and year. Data as in Besley and Reynal-Querol; the full data set
   has 197 distinct countries possibly observed from 1848--2004,
   though few countries have complete data over that period. LOG(GDP
   per capita) is also included in the specification, as are country
   (but not year) dummy variables. Full regression results and
   replication data and Stata code are available at the dataverse for
   this paper}.
 \end{center}
\end{table}
\normalsize

With all data the effect of the democracy dummy on the (linear
probability) of a leader having a graduate degree is 26\% (with a
standard error of 4.3\%; restricting the sample to countries with at
least one leader having a graduate degrees increases this coefficient
to 29.5\% (with a small increase in the standard error); this is a
14\% increase in the estimate coefficient when 17\% of the data are
dropped. The LOGITFE, which automatically drops the ALLZERO groups, show
the average marginal effect of a country being a democracy on either
not having or having a leader with a graduate degree of 29.7\%, almost
exactly the corresponding LPMFE estimated effect \emph{dropping the
  ALLZERO groups}.
For those who doubt the algebra of the previous
  section, I also report the regression results including only the
  ALLZERO groups; the estimated coefficients is, of course, zero (to 17
  decimal places).

\section{Conclusion} \label{s:conc}

The takeaway from this article is fairly simple. Researchers often
require fixed effects specifications to treat unmodeled heterogeneity
which is correlated with the covariates. Such researchers often
either choose LOGIT or OLS without justification, or present the
results of both. While in many cases both LOGIT and OLS yield the
same sign and  crossing of the $p<.05$ level, we have seen that
the appropriate comparison for LOGIT is regression dropping groups
that do not vary on the dependent variable.

One can make a case that either estimate is correct, with the choice
between them being based on theoretical ideas that have no empirical
referent. LPMFE results using all groups, which
average zero with the LPMFE on the restricted data set, make sense in
that the marginal effect of the covariates on $y$ could be thought of
as being zero in the ALLZERO groups. After all, if $P((\ygi=1)=0|g\in
ALL_0$), then then marginal effect of any $x$ in the ALLZERO groups is
indeed zero. Alternatively, we can think of this as a meaningless
exercise, since some change in an $x$ in an ALLZERO group member will
change a failure to a success and thus its marginal effect cannot be
zero. Researchers can report both numbers and their interpretation;
what is clear is that researchers must understand the difference
between the two estimates, and understand how to compare LOGITFE and
LPMFE results. And clearly researchers should not naively compare
OLSFE estimates using all the data with LOGITFE estimates which are
only for the NOTALL0 groups. In any event, the notion that OLSFE is superior to
LOGITFE because the former does not drop the ALLZERO, is clearly
incorrect, and the OLSFE uses the ALLZERO group data in a very
artificial manner unless one \emph{believes} the untestable assumption
that the effect of the
covariates in the ALLZERO groups is really zero.

\section{References}

Beck, Nathaniel. 2018. ``Estimating grouped data models with a binary dependent variable and
  fixed effects: What are the issues?'' http://arxiv.org/abs/1809.06505.

Besley, Timothy \and\ Marta Reynal-Querol. 2011.
``Do Democracies Select More Educated Leaders?'' {\em The American
  Political Science Review} 105:552--566.

Chow, Gregory. 1960.
 ``Tests of Equality Between Sets of Coefficients in Two Linear
  Regressions.'' {\em Econometrica} 28:591--605.

Coup{\'e}, Tom. 2005.
 ``Bias in Conditional and Unconditional Fixed Effects Logit
  Estimation: A Correction.'' {\em Political Analysis} 13:292--295.

Green, Donald, Soo~Yeon Kim \and\ David Yoon. 2001.
 ``Dirty Pool.'' {\em International Organizations} 55:441--68.

Greene, William. 2004.
 ``The behaviour of the maximum likelihood estimator of limited
  dependent variable models in the presence of fixed effects.'' {\em
  Econometrics Journal} 7:98--119.

Greene, William. 2018.
 {\em Econometric Analysis}.
 8 ed. New York, NY:  Pearson.

Katz, Ethan. 2001.
 ``Bias in Conditional and Unconditional Fixed Effects Logit
  Estimation.'' {\em Political Analysis} 9:379--384.

Neyman, J. \and\ E.~L. Scott. 1948.
\ ``Consistent Estimation Based on Partially Consistent Observations.''
  {\em Econometrica} 16:1--32.

Wright, Joseph, Erica Frantz \and\ Barbara Geddes. 2013.
 ``Oil and Autocratic Regime Survival.'' {\em British Journal of
  Political Science} 45:287--306.

\end{document}